\documentclass[preprint,onecolumn,nofootinbib]{revtex4}
\pdfoutput=1
\usepackage{graphicx}
\usepackage{amsmath}
\usepackage{amsfonts}
\usepackage{amssymb}
\usepackage{color}%
\usepackage{dcolumn}
\newcommand{\f}{\begin{equation}}
\newcommand{\ff}{\end{equation}}
\newcommand{\fa}{\begin{eqnarray}}
\newcommand{\ffa}{\end{eqnarray}}

\begin{document}
\title{Holographic Metal-Insulator Transition in Higher Derivative Gravity}
\author{Yi Ling $^{1,3}$}
\thanks{lingy@ihep.ac.cn}
\author{Peng Liu $^{1}$}
\thanks{Corresponding author. liup51@ihep.ac.cn}
\author{Jian-Pin Wu $^{2,3}$}
\thanks{jianpinwu@mail.bnu.edu.cn}
\author{Zhenhua Zhou $^{1}$}
\thanks{zhouzh@ihep.ac.cn}
\affiliation{$^1$ Institute of High Energy Physics, Chinese Academy of Sciences, Beijing 100049, China\ \\
$^2$ Institute of Gravitation and Cosmology, Department of
Physics, School of Mathematics and Physics, Bohai University, Jinzhou 121013, China\ \\
$^3$ Shanghai Key Laboratory of High Temperature Superconductors,
Shanghai, 200444, China}
\begin{abstract}
We introduce a Weyl term into the Einstein-Maxwell-Axion theory in
four dimensional spacetime. Up to the first order of the Weyl
coupling parameter $\gamma$, we construct charged black brane
solutions without translational invariance in a perturbative
manner. Among all the holographic frameworks involving
higher derivative gravity, we are the first to obtain
metal-insulator transitions (MIT) when varying the system
parameters at zero temperature. Furthermore, we study the
holographic entanglement entropy (HEE) of strip geometry in this
model and find that the second order derivative of HEE with
respect to the axion parameter exhibits maximization behavior near
quantum critical points (QCPs) of MIT. It testifies the conjecture
in \cite{Ling:2015dma,Ling:2016wyr} that HEE itself or its
derivatives can be used to diagnose quantum phase transition (QPT).
\end{abstract}
\maketitle
\section{Introduction}
Quantum phase transition (QPT) \cite{Sachdev:2000qpt} occurs at
absolute zero temperature when varying system parameters, which is
believed to account for some peculiar phenomena observed in novel
condensed matter at finite temperature, such as the strange metal.
As a fundamental issue in condensed matter physics, QPT has
attracted a lot of interest of both theorists and experimentalists.
However, QPT usually involves strong correlation physics where
conventional perturbative techniques lose the power.

Recently, gauge/gravity duality
\cite{Maldacena:1997re,Witten:1998qj,Gubser:1998bc} has provided a
novel mechanism to implement QPT in holographic approach,
especially metal-insulator transition (MIT), for instance see
recent review \cite{Ling:2015ghh} and references therein. One key
ingredient of implementing MIT in holography is to introduce the momentum dissipation by
breaking the translational symmetry, meanwhile deform
the near horizon geometry to new IR fixed point which is dual to
an insulating phase \cite{Donos:2012js,Donos:2013eha}. A
simple holographic model with momentum dissipation can be
constructed by introducing a set of massless axion fields
\cite{Andrade:2013gsa}. In its original version, only the metallic
phase was found. Later, when a general potential of axions or an
additional dilaton field is introduced into this
Einstein-Maxwell-Axion (EMA) model, the insulating phase has also
been observed
\cite{Baggioli:2014roa,Baggioli:2016oqk,Baggioli:2016oju,Kiritsis:2015oxa,Gouteraux:2014hca,Gouteraux:2016wxj}.
Recent investigation on the hydrodynamic and transport
properties of EMA model without translational invariance can be found, for instance in
\cite{Ling:2016ien,Cheng:2014tya,Wang:2016vmm}.

In this paper, we provide a new strategy to implement MIT by
introducing higher-derivative terms into the EMA model. This is
also the first time realizing MIT in the framework of higher
derivative gravity. In four dimensional spacetime, there are eight
independent terms in a general four-derivative action
\cite{Myers:2010pk}, which may emerge as quantum corrections in
the low energy effective action of superstring theory
\cite{Hanaki:2006pj,Cremonini:2008tw}. From the viewpoint of the
dual conformal field theory (CFT), these terms correspond to the
corrections of finite `t Hooft coupling and/or beyond the
large-$N$ limit. Here, we only focus on a special term with
coupling between gauge field and the Weyl tensor, which has been
dubbed as Weyl term \cite{Myers:2010pk,Myers:2009ij,Ritz:2008kh}.
In holographic literature, it has been shown that the presence of
the Weyl term in four-dimensional Schwarzschild-AdS geometry
induces non-trivial behavior of the conductivity in dual theory
\cite{Myers:2010pk}, in contrast to the frequency-independent
conductivity without Weyl term \cite{Herzog:2007ij}. Specifically,
the real part of the conductivity displays a peak or a valley near
zero frequency which depends on the sign of Weyl coupling
parameter, implying that the frequency-dependent conductivity is
particle-like or vortex-like, respectively \cite{Myers:2010pk}
(also see
\cite{WitczakKrempa:2012gn,WitczakKrempa:2013ht,Witczak-Krempa:2013nua,Witczak-Krempa:2013aea,Myers:2016wsu}).

Once the Weyl term is taken into account, the equations of motion
for this Einstein-Maxwell-Axion-Weyl (EMA-Weyl) theory become a
set of third order differential equations with high nonlinearity,
which is very hard to solve analytically so far. As the first
step, we treat the Weyl coupling parameter $\gamma$ as a small
number and construct analytical solutions up to the first order of
this parameter. This strategy has previously been used to
construct perturbative charged black hole solutions for
high-derivative gravity in
\cite{Liu:2008kt,Myers:2009ij,Cai:2011uh,Dey:2015poa,Dey:2015ytd,Mahapatra:2016dae}.
With the background solutions up to $\mathcal O(\gamma)$, we will
study the thermodynamics of the dual field theory. Moreover, the
direct current (DC) conductivity can be derived analytically,
which allows us to study the phase structure at zero temperature
directly. We will demonstrate that MIT as a quantum critical
phenomenon can be observed in this circumstance manifestly.

Stimulated by our recent work \cite{Ling:2016wyr,Ling:2015dma},
we intend to investigate the behavior of holographic entanglement
entropy (HEE) close to QPT in this holographic model. In condensed
matter physics, a lot of work have revealed that  the entanglement
itself or its derivatives displays local extremes close to QCPs
\cite{Amico:2007ag,Chen:2006eqp,Hamma:2007eft,Osterloh:2002na,Osborne:2002zz,
Vidal:2002rm,YChen:2006jop,Anfossi:2006tvm,Wu:2006leq,Wu:2004prl,Larsson2005eso}.
Nevertheless, this phenomenon calls for deeper theoretical
understanding. Recently our series of work
\cite{Ling:2015dma,Ling:2016wyr} have disclosed that the HEE or
its first order derivatives with respect to system parameters can
be used to characterize the QPT such that a holographic
description of the relation between EE and QPT has been
established. We have also proposed that it should be a universal
feature that the HEE or its derivatives with respect to system
parameters can diagnose  the QPT in a generic holographic
framework. The robustness of this proposal awaits for further test.
Therefore, it is very natural to ask whether the quantum
critical phenomena observed in this model can also be captured by
HEE. Interestingly enough, in this paper we
will demonstrate that the \emph{second} order derivative of HEE does
exhibit maximization behavior close to QCPs. This observation not
only justifies the conjecture in \cite{Ling:2015dma,Ling:2016wyr},
but also enriches our understanding on the scenario of HEE
characterizing QPT.

Our paper is organized as follows. We shall firstly introduce a
Weyl term into four dimensional EMA theory and then obtain the
perturbative black hole solutions in Section \ref{BH}, with a
brief discussion on the thermodynamics of the background. In
Section \ref{MIT}, we calculate DC conductivity of the dual system
at absolute zero temperature and then demonstrate that MIT takes
place as quantum critical dynamics. Then we move on to the study
of the HEE in Section \ref{HEE} and show that the second order
derivative of HEE with respect to system parameter exhibits peaks
close to QPTs. In Section \ref{Con}, we summarize the results of
this paper and discuss some open questions for further investigation.

\section{Einstein-Maxwell-axions-Weyl model}\label{BH}

\subsection{Setup and equations of motion}

We consider a four dimensional EMA theory with an additional Weyl term coupled to the Maxwell field
(EMA-Weyl model), whose action reads as
\begin{eqnarray}
\label{action}
S=\int d^4x\sqrt{-g}\left[R+6-\frac{1}{4}F_{\mu\nu}F^{\mu\nu}-\sum_{I=x,y}(\partial \phi_I)^2+\gamma C_{\mu\nu\rho\sigma}F^{\mu\nu}F^{\rho\sigma}\right]
\,,
\end{eqnarray}
where $F=dA$ and $\gamma$ is the Weyl coupling parameter.
$\phi_I$ is a set of free massless axion fields, responsible
for the momentum dissipation \cite{Andrade:2013gsa}. $C_{\mu\nu\rho\sigma}$ is the Weyl tensor,
which is defined as
\begin{eqnarray}
\label{WT}
C_{\mu\nu\rho\sigma}=R_{\mu\nu\rho\sigma}-(g_{\mu[\rho}R_{\sigma]\nu}-g_{\nu[\rho}R_{\sigma]\mu})+\frac{1}{3}R
g_{\mu[\rho}g_{\sigma]\nu}.
\end{eqnarray}

It is straightforward to derive the equations of motion from the action in (\ref{action}), which read as
\begin{eqnarray}
&&
\Box\phi_I=0
\,,
\label{phiE}
\\
&&
\nabla_{\mu}[F^{\mu\nu}-4\gamma  C^{\mu\nu\rho\sigma}F_{\rho\sigma}]=0
\,,
\label{ME}
\\
&&
R_{\mu\nu}-\frac{1}{2}R g_{\mu\nu}-3g_{\mu\nu}
-\frac{1}{2}\Big(F_{\mu\rho}F_{\nu}^{\ \rho}-\frac{1}{4}g_{\mu\nu}F_{\rho\sigma}F^{\rho\sigma}\Big)
-\partial_\mu\phi_x\partial_\nu\phi_x+\frac{g_{\mu\nu}}{2}(\partial \phi_x)^2
\nonumber
\\
&&
-\partial_\mu\phi_y\partial_\nu\phi_y+\frac{g_{\mu\nu}}{2}(\partial \phi_y)^2
-\gamma(G_{1\mu\nu}+G_{2\mu\nu}+G_{3\mu\nu})
=0\,,
\label{EE}
\end{eqnarray}
where
\begin{eqnarray}
\label{G1}
G_{1\mu\nu}&=&\frac{1}{2}g_{\mu\nu}R_{\alpha\beta\rho\sigma}F^{\alpha\beta}F^{\rho\sigma}
-3R_{(\mu|\alpha\beta\lambda|}F_{\nu)}^{\ \alpha}F^{\beta\lambda}
-2\nabla_{\alpha}\nabla_{\beta}(F^{\alpha}_{\ (\nu}F^{\beta}_{\ \mu)})
\,,
\
\\
\nonumber
G_{2\mu\nu}&=&-g_{\mu\nu}R_{\alpha\beta}F^{\alpha\lambda}F^{\beta}_{\ \lambda}
+g_{\mu\nu}\nabla_{\alpha}\nabla_{\beta}(F^{\alpha}_{\ \lambda}F^{\beta\lambda})
+\Box(F_{\mu}^{\ \lambda}F_{\nu\lambda})
-2\nabla_\alpha\nabla_{(\mu}(F_{\nu)\beta}F^{\alpha\beta})
\
\\
\label{G2}
&&
+2R_{\nu\alpha}F_{\mu}^{\ \beta}F^{\alpha}_{\ \beta}
+2R_{\alpha\beta}F^{\alpha}_{\ \mu}F^{\beta}_{\ \nu}
+2R_{\alpha\mu}F^{\alpha\beta}F_{\nu\beta}
\,,
\
\\
\label{G3}
G_{3\mu\nu}&=&
\frac{1}{6}g_{\mu\nu}RF^2-\frac{1}{3}R_{\mu\nu}F^2-\frac{2}{3}RF^{\alpha}_{\ \mu}F_{\alpha\nu}
+\frac{1}{3}\nabla_{(\nu}\nabla_{\mu)} F^2-\frac{1}{3}g_{\mu\nu}\Box F^2\,.
\end{eqnarray}

As pointed out in the introduction, it is difficult to solve
this system with full backreaction. Next we intend to construct a
perturbative charged black brane solution to this system up to
$\mathcal O(\gamma)$.

\subsection{Charged black brane solutions}

In this subsection, we intend to construct a charged black brane
solutions by solving above equations (\ref{phiE}),
(\ref{ME}) and (\ref{EE}). To this end, we take the following
ansatz,
\begin{eqnarray}
ds^2&=&-r^2f(r)dt^2+\frac{1}{r^2f(r)}dr^2+r^2g(r)(dx^2+dy^2)\,,
\nonumber
\\
A&=&A_t(r)dt\,,~~~\phi_x=kx\,,~~~\phi_y=ky
\,,
\label{ds}
\end{eqnarray}
where the UV boundary is located at $r\rightarrow\infty$.
A non-zero $A_t(r)$ is introduced for a finite chemical potential.
The special form of $\phi_I$ in (\ref{ds}) retains homogeneity as well as isotropy of spacetime
but dissipates the momentum of the UV boundary CFT.
$k$ is the system parameter and also is referred to as axionic charge.
This model does not have manifest lattice wave vector
but captures features of disorder, which is characterized by the
axionic charge $k$ \cite{Grozdanov:2015qia,Gouteraux:2016wxj,Baggioli:2016oqk}.

Since the equations of motion for the axion fields $\phi_I$
in (\ref{phiE}) are not influenced by the corrections from Weyl
term, we only need to expand the functions $f(r)$, $g(r)$
and $A_t(r)$ in powers of $\gamma$ up to the first order,
\begin{eqnarray}
&&
\nonumber
f(r)=f_0(r)+\gamma Y(r)
\,,
\
\\
&&
\nonumber
g(r)=1+\gamma G(r)
\,,
\
\\
&&
A_{t}(r)=A_{t0}(r)+\gamma H(r)
\,,
\end{eqnarray}
where $f_0(r)$ and $A_{t0}(r)$ are leading order solutions,
while $G(r)$, $H(r)$ and $Y(r)$ are corrections of order $\mathcal O(\gamma)$.

These functions can be determined by directly solving the
equations of motion (\ref{phiE}), (\ref{ME}) and (\ref{EE}) to
the zeroth and first order of $\gamma$,
\begin{eqnarray}
&&
f_0(r)=1-\frac{M}{r^3}+\frac{q^2}{r^4}-\frac{k^2}{r^2}\,,
\label{f0}
\\
&&
A_{t0}(r)=\mu-\frac{2q}{r} \,,
\label{At0}
\\
&&
G(r)=\frac{4 q^2}{9 r^4}-\frac{g_0}{r}+g_1 \,,
\label{Gr}
\\
&&
H(r)=\frac{296 q^3}{45 r^5}-\frac{4 M q}{r^4}-\frac{16 k^2 q}{9 r^3}-\frac{g_0 q}{r^2} \,,
\label{Hr}
\\
&&
Y(r)=-\frac{104 q^4}{45 r^8}+\frac{20 M q^2}{9 r^7}-\frac{32 q^2}{9 r^4}+\frac{20 k^2 q^2}{9 r^6}
+\frac{g_0 q^2}{r^5}-\frac{g_0 M}{2 r^4}+\frac{g_1 k^2}{r^2}-\frac{g_0}{r}\,.
\label{Yr}
\end{eqnarray}
Eqs. (\ref{f0}) and (\ref{At0}) are exactly the solutions of EMA
model proposed in \cite{Andrade:2013gsa}. In above
equations, there are five integration constants
$(\mu,q,M,g_0,g_1)$, which are not independent from one
another. Subsequently, we shall derive the relations among these
parameters.

First, one can show that the integration constants $g_0$ and
$g_1$ can be eliminated by coordinate transformations
$r\rightarrow r+\gamma g_0/2,~x\rightarrow x(1-\gamma
g_1/2),~y\rightarrow y(1-\gamma g_1/2)$ as well as a redefinition
of the axion charge $k\to k(1+\gamma g_1/2$). Then, up to
$\mathcal O(\gamma)$ Eqs. (\ref{f0} - \ref{Yr}) can be reexpressed
as
\begin{eqnarray}
&&
f_0(r)=1-\frac{M}{r^3}+\frac{q^2}{r^4}-\frac{k^2}{r^2}\,,
\nonumber
\\
&&
A_{t0}(r)=\mu-\frac{2q}{r} \,,
\nonumber
\\
&&
G(r)=\frac{4 q^2}{9 r^4} \,,
\nonumber
\\
&&
H(r)=\frac{296 q^3}{45 r^5}-\frac{4 M q}{r^4}-\frac{16 k^2 q}{9 r^3} \,,
\nonumber
\\
&&
Y(r)=-\frac{104 q^4}{45 r^8}+\frac{20 M q^2}{9 r^7}-\frac{32 q^2}{9 r^4}+\frac{20 k^2 q^2}{9 r^6}\,.
\label{fr2}
\end{eqnarray}
Furthermore, the location of the horizon $r_h$ of this Weyl corrected charged black brane solutions is determined by
\begin{eqnarray}
\label{frh}
f(r_h)=0\,.
\end{eqnarray}
In the mean time, to guarantee that $A$ is well-defined on the horizon, we need to set
\begin{eqnarray}
\label{Atrh}
A_t(r_h)=0\,.
\end{eqnarray}
We can refer above two conditions as the horizon conditions, which
give the relations among $(\mu,q,M)$ as
\begin{eqnarray}
&&
\label{qbymu}
q=\frac{r_h \mu }{2}+\gamma\Big(\frac{29 \mu ^3}{180 r_h}-r_h \mu+\frac{5 k^2 \mu }{9 r_h} \Big)\,,
\\
&&
\label{Mbymu}
M=\frac{r_{h }\mu ^2}{4}+r_h^3-k^2 r_h+\gamma\Big(\frac{7 \mu ^4}{45 r_h}-\frac{4r_{h } \mu ^2}{3}+\frac{5 k^2 \mu ^2}{9 r_h}\Big)\,.
\end{eqnarray}
And then it is straightforward to derive the Hawking temperature of this black brane as
\begin{eqnarray}
\label{HTem}
T=-\frac{4 k^2+\mu ^2-12 r_h^2}{16 \left(\pi  r_h\right)}+\gamma\frac{\left(\mu ^4-60 \mu ^2 r_h^2\right) }{720 \pi  r_h^3} \,.
\end{eqnarray}
Note that we have expanded all the above quantities $q$, $M$ and $T$ to $\mathcal O(\gamma)$.

\subsection{Thermodynamics}

In this subsection, we briefly discuss the thermodynamics of the quantum field theory dual to the
EMA-Weyl system with a standard approach (see e.g. \cite{Hartnoll:2009sz}). To this end, we first construct the renormalized action
$S_{ren}$ by adding a boundary term to the original action (\ref{action}) as
\begin{eqnarray}
 S_{ren}=S+S_{bdy}=S+\int_{r\rightarrow\infty} dx^3 \sqrt{h}(2\mathcal{K}-4)\,,
\end{eqnarray}
where $h$ is the determinant of the induced metric $h_{ij}$ on the boundary and $\mathcal{K}$ is the trace of the extrinsic curvature on slice with constant $r$.
By a straightforward calculation, the free energy density can be derived as
\begin{eqnarray}
F=\left(-\frac{1}{4} \mu ^2 r_h-r_h^3-k^2 r_h\right)+\gamma\left(-\frac{7 \mu ^4}{45 r_h}+\frac{4 \mu ^2 r_h}{3}-\frac{5 k^2\mu^2}{9r_h}\right) \,.
\end{eqnarray}

And then, we derive the charge density $Q$, entropy density $s$, pressure $P$ and energy density $\epsilon$ as
\begin{eqnarray}
&&Q=-\frac{\partial F}{\partial \mu}=\mu  r_h+\frac{\gamma  \left(-180 \mu  r_h^2+100 k^2 \mu +29 \mu ^3\right)}{90 r_h}\,,\\
&&s=-\frac{\partial F}{\partial T}=4 \pi  r_h^2-\frac{20}{9} \gamma  \left(\pi  \mu ^2\right)\,,\\
&&P=-F=\left(\frac{1}{4} \mu ^2 r_h+r_h^3+k^2 r_h\right)+\gamma\left(\frac{7 \mu ^4}{45 r_h}-\frac{4 \mu ^2 r_h}{3}+\frac{5 k^2\mu^2}{9r_h}\right) \,,\\
&&\epsilon=2 r_h^3+\frac{\mu ^2 r_h}{2}-2 k^2 r_h+\frac{\gamma  \left(-120 \mu ^2 r_h^2+50 k^2 \mu ^2+14 \mu ^4\right)}{45 r_h}\,.
\end{eqnarray}
where $\epsilon=s T+\mu Q-P$ \cite{Hartnoll:2009sz} has been used in the last equation.
Again, all above thermodynamical quantities are obtained up to $\mathcal O(\gamma)$.

One can also check that  $s$ and $\epsilon$ are the Wald entropy density \cite{Wald:1993nt} and ADM  mass density respectively \cite{Ashtekar:1984,Ashtekar:1999jx}.
In addition, for vanishing axions, \emph{i.e.}, $k=0$, our results agree with that in  \cite{Dey:2015poa,Dey:2015ytd}.

\section{Metal-insulator transition}\label{MIT}

In this section we study the MIT by analyzing the behavior
of direct current (DC) conductivity. By definition, a MIT is
reflected by an abrupt change of DC conductivity behavior, which
is a macroscopic observable governed by quantum critical physics.
To be more specific, at zero temperature, the DC conductivity of a
metallic phase behaves as $\partial_{T}\sigma_{DC}<0 $, while for
insulating phase, it behaves as $\partial_{T}\sigma_{DC}>0$. Then
the critical point (line) of MIT is determined by
$\partial_{T}\sigma_{DC}=0$. Therefore, to demonstrate the MIT
from EMA-Weyl model, we shall firstly calculate the DC
conductivity in what follows.

In the calculation of DC conductivity, it is more readily adapted to work in coordinate $z\equiv r_h/r$.
Note that for a given $\gamma$, the Weyl corrected EMA black hole solutions are parametrized by two scaling-invariant parameters,
$\hat{k}\equiv k/{\mu}$ and $\widehat{T}\equiv T/{\mu}$. With these in mind, we rewrite the background solutions as follows
\begin{eqnarray}
ds^2&=&
\frac{1}{z^2}\Big[-f(z)dt^2+\frac{1}{f(z)}dz^2+g(z)(dx^2+dy^2)\Big]\,,
\nonumber
\\
A&=&A_t(z)dt\,,
\label{dsz}
\end{eqnarray}
where $f(z)$, $g(z)$ and $A_{t}(z)$ now take the form as
\begin{equation}
\begin{aligned}
f(z)=&\,(1-z)p(z)\,,\qquad g(z)=1+\gamma\frac{\mu^2 z^4}{9},\\
p(z)=&\,\gamma\frac{\mu^2z^3}{180}[240+100\hat{k}^2\mu^2(z^3-1)+2\mu^2(13z^4-14)+(\mu^2-100)(z^3+z^2+z)]\\
& +1+z+(1-\hat{k}^2\mu^2)z^2-\frac{\mu^2}{4}z^3,\\
A_t(z)=\,&\mu\Big[(1-z)+\gamma\frac{z}{90}\big(180+20\hat{k}^2\mu^2(-5-4z^2+9z^3)-29\mu^2+74\mu^2z^4-45z^3(4+\mu^2)\big)\Big]\,.
\end{aligned}
\end{equation}
Also the dimensionless Hawking temperature $\hat{T}\equiv T/\mu$, which is given by
\fa
\hat{T}=\frac{12-\mu^2-4\hat{k}^2\mu^2}{16\pi\mu}+\gamma\frac{\mu(\mu^2-60)}{720\pi}\,.
\label{temh}
\ffa

Now, we calculate the DC conductivity in dual field theory employing  the scheme proposed in \cite{Donos:2014uba} (also see \cite{Blake:2014yla}).
We turn on a constant electric field from the beginning, instead of an alternating current (AC) electric field. Specifically, we take the following consistent ansatz
\begin{align}
\delta A_x=-E_xt+a_x(z)\,,~~~\delta g_{tx}=\frac{1}{z^2}[h_{tx}(z)+\gamma G(z)h_{tx}(z)]\,,~~~
\delta\phi_x=\chi_x(z)\,.
\end{align}
The key point of this method is
to find the conserved current in the bulk \cite{Donos:2014uba,Blake:2014yla}, which is in this model
\fa
J^x=\sqrt{-g}(F^{zx}-4\gamma C^{zx\alpha\beta}F_{\alpha\beta})\,.
\label{Jx}
\ffa
Up to $\mathcal O(\gamma)$, $J^x$ can be expressed as
\begin{align}
J^x= -Q h_{tx}+f a'_x-\frac{2}{3}\gamma z^2 f a'_x f''\,.
\end{align}
Here we have denoted $J^t=Q$, which is the conserved electric charge density.
The DC conductivity can then be obtained from the following expression
\fa
\label{sigmaDC}
\sigma_{DC}=\frac{J^{x}}{E_x}\,.
\label{DC1}
\ffa

As revealed in \cite{Donos:2014uba,Blake:2014yla}, given a conserved current $J^x$ along the radial direction,
it is enough to determine the DC conductivity from the requirement of the regularity of the perturbation variables at the horizon $z=1$.
We illuminate this procedure in what follows.

First, to have a well defined gauge field at the horizon, we have
\fa
a'_x(z)=\frac{E_x}{f(z)}\,.
\label{axbeh}
\ffa
Second, when the momentum conservation is violated, $h_{tx}$ should be finite at the horizon and we can extract this value
from the $t-x$ component of Einstein equation, which reads after taking value at $z=1$
\fa
h_{tx} \Big[6 (-6+2 k^2 + f' (-4+\gamma G')+ f'')
+ A^{\prime2}_t (-3+8\gamma f'-4 \gamma f'')\Big]
-2 f a'_x A'_t (3+4\gamma f'')=0\,.
\nonumber
\\
\label{htxbeh}
\ffa
And finally, combining
Eqs. (\ref{Jx}), (\ref{axbeh}) and (\ref{htxbeh}),
the DC conductivity can be expressed as a function of $(\hat k,\mu,\gamma)$,
\fa
\sigma=1+\frac{1}{2\hat{k}^2}+\gamma\Big(4-\frac{8}{3}\hat{k}^2\mu^2+\frac{\mu^2}{9}+\frac{\frac{4}{15}\mu^2-2}{\hat{k}^2}\Big)\,.
\label{DC2}
\ffa

\begin{figure}
\center{
\includegraphics[scale=0.6]{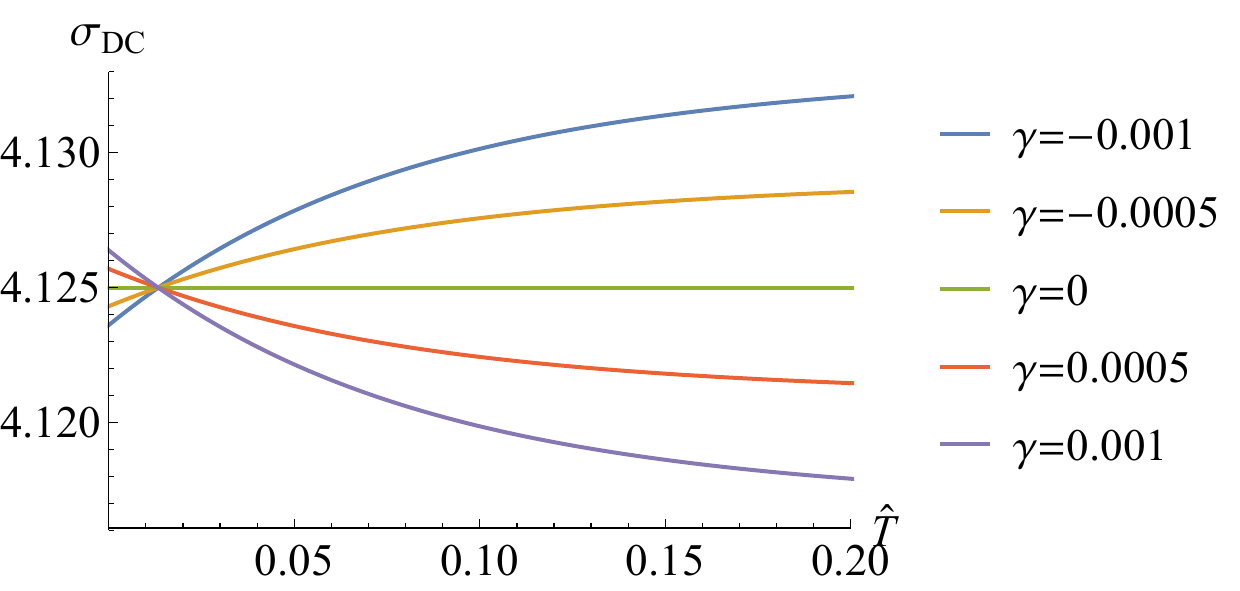}\ \hspace{0.5cm}
\includegraphics[scale=0.6]{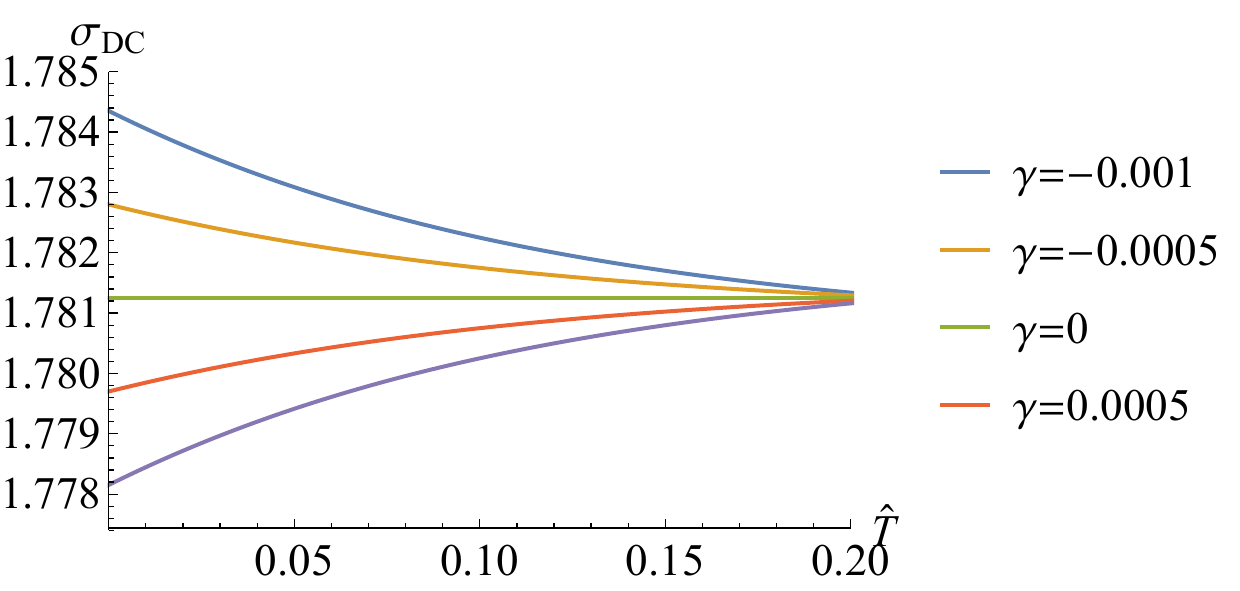}\ \\
\caption{\label{svst} The DC conductivity $\sigma_{DC}$ as the function of the temperature $\hat{T}$
for some specific $\hat{k}$ (left plot for $\hat{k}=0.4$ and right plot for $\hat{k}=0.8$) and $\gamma$.}}
\end{figure}

\begin{figure}
\center{
\includegraphics[scale=0.65]{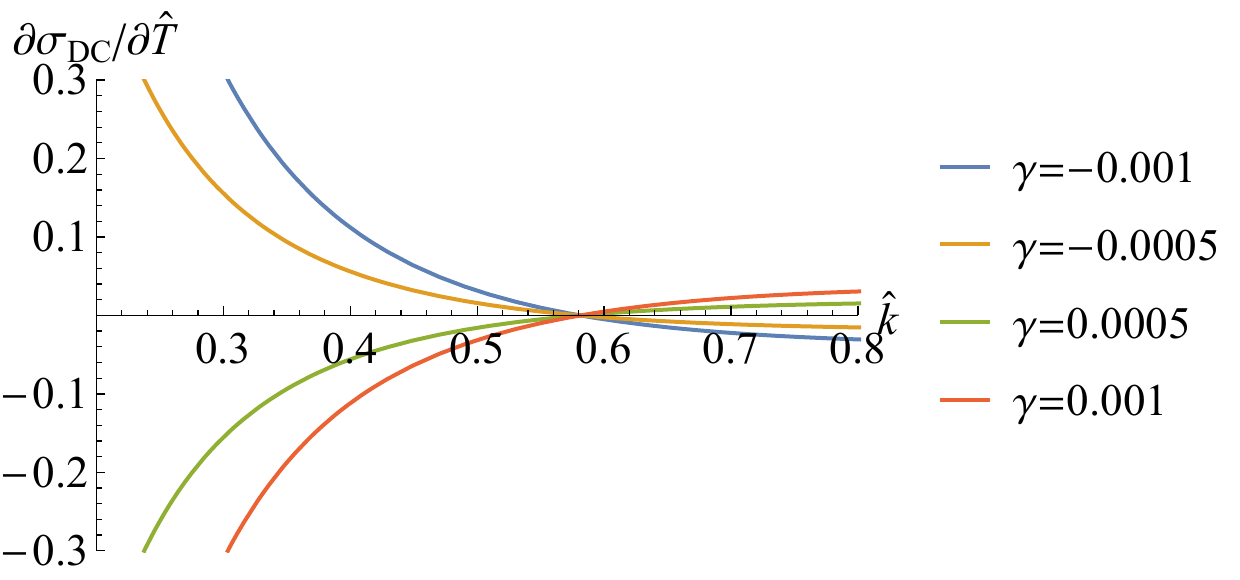}\ \hspace{1.2cm}
\includegraphics[scale=0.44]{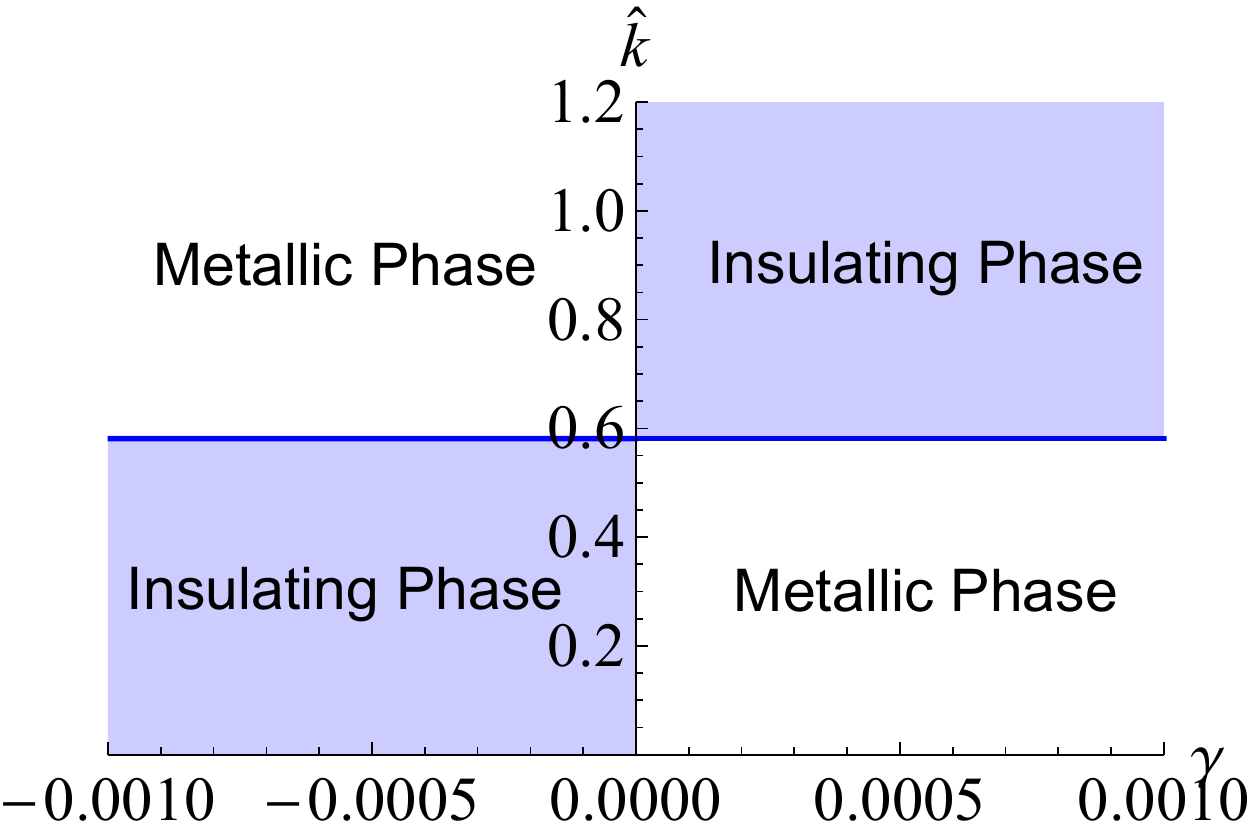}\ \\
\caption{\label{PDtem0hee}
Left plot: $\partial_{\hat{T}}\sigma_{DC}$ as a function of $\hat{k}$ at zero temperature for different Weyl parameters $\gamma$.
Right plot:
The phase diagram over $(\gamma,\hat{k})$ plane for the MIT at zero temperature in the Weyl corrected EMA geometry.
The transverse blue line corresponds to the critical line $\hat{k}\simeq 0.58116$.
Note that $\gamma\neq 0$ here.}}
\end{figure}

Armed with (\ref{DC2}), we study the MIT by examining the behavior
of $\sigma$ in zero temperature limit. First, we directly see from
(\ref{DC2}) that when $\gamma=0$, the DC conductivity is
independent of the temperature, which has been observed in
\cite{Andrade:2013gsa}. But when $\gamma\neq 0$, the DC
conductivity is temperature dependent (Fig. \ref{svst}). In
particular, we find that given a nonzero $\gamma$,
a MIT occurs when varying the system parameter
$\hat{k}$ (see the left plot in Fig.\ref{PDtem0hee}). To
demonstrate the MIT of this model more explicitly, we plot
the phase diagram over $(\gamma,\hat{k})$ plane at zero
temperature in Fig.\ref{PDtem0hee} (right plot). The
quantum critical line (blue line in the right plot in
Fig.\ref{PDtem0hee}) is determined by
$\partial_{\hat{T}}\sigma_{DC}=0$ at zero temperature, as is
discussed at the beginning of this section, which corresponds to
$\hat{k}=\frac{1}{4}\sqrt{\frac{1}{15}(5+\sqrt{5785})}\simeq
0.58116$. This quantum critical line is independent of $\gamma$.
Specifically, we observe that for $\gamma>0$, the transition
from metallic phase to insulating phase occurs when increasing
$\hat k$. For $\gamma<0$, however, the opposite scenario is
obtained.


In
\cite{Myers:2010pk,WitczakKrempa:2012gn,WitczakKrempa:2013ht,Witczak-Krempa:2013nua,Witczak-Krempa:2013aea,Myers:2016wsu},
the transport behavior of the boundary field theory dual to Schwarzschild-AdS geometry has been studied.
The optical conductivity near the zero frequency displays a Drude-like peak
for $\gamma>0$, while for $\gamma<0$ the conductivity exhibits a
valley near zero frequency. As argued in \cite{Myers:2010pk} (also
see
\cite{WitczakKrempa:2012gn,WitczakKrempa:2013ht,Witczak-Krempa:2013nua,Witczak-Krempa:2013aea,Myers:2016wsu}),
the Drude-like behavior for $\gamma>0$ could be described by the
collision and motion of charged particles, while the valley
behavior for $\gamma<0$ should be depicted by the collision of
vortices. Further, it is revealed in
\cite{Myers:2010pk,WitczakKrempa:2012gn,WitczakKrempa:2013ht,Witczak-Krempa:2013nua,Witczak-Krempa:2013aea,Myers:2016wsu}
that for small $\gamma$, an EM duality-transformation relates the
equations of motions of the Maxwell field at $\gamma$ and the one
at $-\gamma$, leading to ${\sigma}(\omega,\gamma)\simeq
1/\sigma(\omega,-\gamma)$ in the dual boundary theory. Later, a
broader class of particle-vortex duality was revealed in
\cite{Burgess:2000kj,Murugan:2016zal}, which is different
from that in
\cite{Myers:2010pk,WitczakKrempa:2012gn,WitczakKrempa:2013ht,Witczak-Krempa:2013nua,Witczak-Krempa:2013aea,Myers:2016wsu}.
From Fig. \ref{svst} we find that in our model
$\sigma_{DC}(\gamma,T)$ exhibits an interesting mirror
symmetry
\begin{equation}\label{pvd1}
  \sigma_{DC}(\gamma,T)\simeq const. - \sigma_{DC}(-\gamma,T),
\end{equation}
when $\hat k$ is fixed, which can be viewed as a special
particle-vortex duality as investigated in
\cite{Burgess:2000kj,Murugan:2016zal}. It can be deduced
from Eq.(\ref{pvd1}) that $\partial_T \sigma(\gamma,T)$ is
an odd function of $\gamma$, which is also numerically
depicted in the left plot of Fig.\ref{PDtem0hee}. Thus, we
have a ``metal-insulator'' duality when changing the sign of
$\gamma$, as illustrated in the right plot of
Fig.\ref{PDtem0hee}. A concrete and analytical derivation
of Eq.(\ref{pvd1}) in our present model, however, would be
more complicated and difficult than that in
\cite{Myers:2010pk,WitczakKrempa:2012gn,WitczakKrempa:2013ht,Witczak-Krempa:2013nua,Witczak-Krempa:2013aea,Myers:2016wsu}
due to the involvement of finite charge density and
momentum dissipation. We leave this issue for future investigation.

Our present holographic EMA-Weyl model is dual to a boundary field theory
with finite density and momentum dissipation. For weak momentum
dissipation (small $k$), the transport behaves as metallic for
$\gamma>0$, which may be described by the motion and
collision of particles \cite{Myers:2010pk}. With the
increase of $\hat k$ the motion of particles is suppressed and the
system undergoes a phase transition from a metallic phase
to an insulating phase. It is deduced from Eq.(\ref{pvd1})
that an opposite scenario happens for $\gamma<0$, which results in
the phase structure as illustrated in the right plot of
Fig. \ref{PDtem0hee}.

\section{Holographic entanglement entropy close to QCPs}\label{HEE}

In this section, we study the HEE for the dual field theory living on the boundary.
It has been revealed in \cite{Ling:2015dma,Ling:2016wyr} that HEE or its first order derivative with
respect to system parameters exhibit local extremes near QCPs, and thus can be used to diagnose QPT in
holographic framework. It is also conjectured in \cite{Ling:2016wyr} that higher-order derivatives of HEE
probably play a similar role in  characterizing QPT in holographic models. Inspired by this observation, we
intend to compute the HEE in our present model. In comparison with the previous holographic models with
MIT, one nice feature of our current model is that  analytical solutions for black  brane  with zero temperature
are derived such that we can directly compute the HEE in a bulk geometry at zero temperature.

Before calculating the HEE explicitly, it is worthwhile to point out a key difference in the holographic
description of EE in higher derivative gravity. It is noticed that the original Ryu-Takayanagi
formula \cite{Ryu:2006bv,Ryu:2006ef} only holds for Einstein gravity. In \cite{Hung:2011xb}, an
alternative prescription for HEE is proposed for Lovelock gravity.
This prescription reproduces the universal contribution to the EE for the dual CFT in four and six dimensional spacetimes.
Further, a general formula for HEE in higher derivative gravity is proposed in \cite{Dong:2013qoa}.
This formula includes the Wald entropy as the leading term and a correction from extrinsic curvature,
which is usually dubbed as the anomaly term of HEE. For our model, it is easy to check that the
anomaly term of HEE vanishes for the Weyl corrected action (\ref{action}). Therefore, on a slice
with fixed time, it is valid to calculate the HEE of our present model with the leading term of the
formula proposed in \cite{Dong:2013qoa}, which is
\fa
\label{HEEWeyl}
S_{EE}=-2\pi\int_\Sigma d^2x\sqrt{h}\frac{\partial \mathcal{L}}{\partial R_{\mu\nu\rho\sigma}}\varepsilon_{\mu\nu}\varepsilon_{\rho\sigma}\,,
\ffa
where $\mathcal{L}$ is the Lagrangian density of action (\ref{action}), $\varepsilon_{\mu\nu}$ is the Levi-Civita symbol,
and $h$ is determinant of the induced metric on the surface $\Sigma$ that minimizes the functional $S_{EE}$.
In this EMA-Weyl model, the formula (\ref{HEEWeyl}) can be evaluated as
\fa
S_{EE}=4\pi\int_\Sigma d^2x \sqrt{h} \Big(1+\frac{1}{3}\gamma F^2\Big).
\ffa

Now we compute the HEE in a bulk geometry at zero
temperature. We consider a strip geometry on the dual boundary
system that has length $L_y\to \infty$ in $y$-direction and finite width
$\hat l$ in $x$-direction. Given the fact that both $h$ and $F$
are functions of radial axis $z$ only, we can label the
$\Sigma$ with the location of its bottom $z_*$ in $z$-direction.
Since the boundary is asymptotically $AdS_4$, the $S_{EE}$ for
each background solution will receive a vacuum contribution.
Here we define the HEE as $\hat{S}\equiv \left(S_{EE} -
S_{vac}\right)/2\pi L_y$ with vacuum contribution $S_{vac}$ subtracted
out. The scaling-invariant width $l$ and HEE $S$ can be expressed
as,
\begin{eqnarray}\label{hee}
S &=& \frac{4}{\mu}\left(\frac{1}{z_*} +\int_{0}^{z_*}dz\left[\frac{{{{\left( {\gamma {F^2}(z) + 3} \right)}^2}
{g_{{yy}}}(z)\sqrt {{g_{{xx}}}(z){g_{zz}}\left( z \right)} }}{{3\sqrt \xi  }} - \frac{1}{{{z^2}}}\right]\right),\\
  l &=&  2\mu\int_{0}^{z_*}dz\left( {\gamma {F^2}\left( {{z_*}} \right) + 3} \right)\sqrt {\frac{{{g_{{xx}}}
  \left( {{z_*}} \right){g_{{yy}}}\left( {{z_*}} \right){g_{{zz}}}(z)}}{{{g_{{xx}}}(z)\xi }}},
\end{eqnarray}
where $l=\mu\hat l,\, S=\hat S/\mu$, and $\xi  \equiv {\left( {\gamma {F^2}(z) + 3} \right)^2}{g_{{xx}}}(z){g_{{yy}}}(z) - {\left( {\gamma {F^2}\left( {{z_*}} \right) + 3} \right)^2}{g_{{xx}}}\left( {{z_*}} \right){g_{{yy}}}\left( {{z_*}} \right)$.

\begin{figure}
  \centering
  \includegraphics[width=10cm]{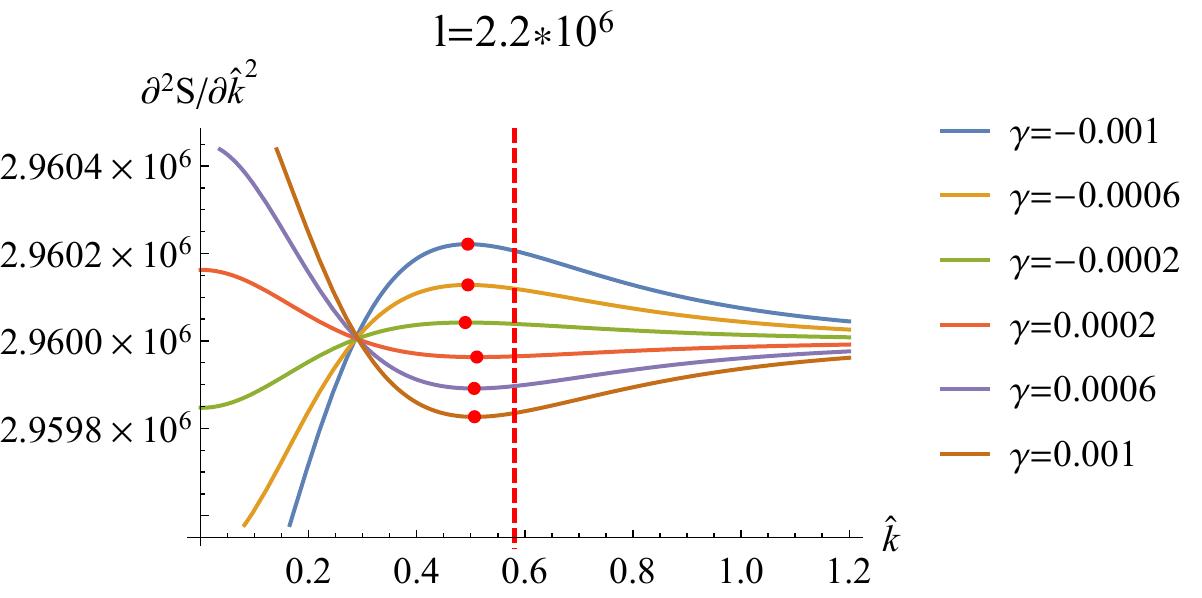}\\
  \caption{Each curve represents the $\partial^2 S/\partial \hat k^2 \; v.s.\; \hat k$ with $\gamma$ specified by the plot legends.
  The red dashed line is $\hat k_c = 0.58116$.}\label{line1}
\end{figure}
\begin{figure}
  \centering
  \includegraphics[width=10cm]{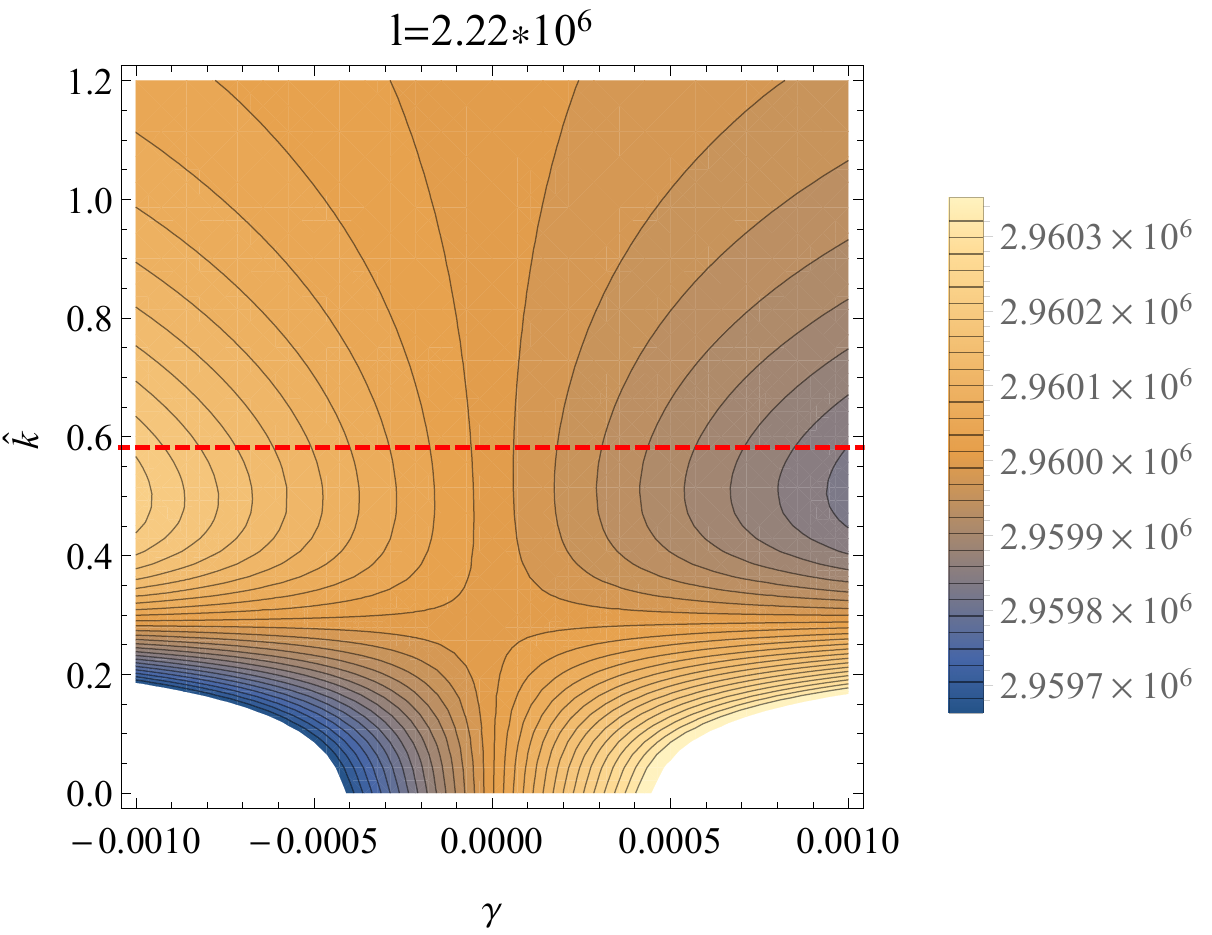}\\
  \caption{The contour plot of $\partial^2 S/\partial \hat{k}^2$ over $(\gamma,\hat{k})$ at $l=2.22 \times 10^6$ and $\hat{T}=0$.
  The values of $\partial^2 S/\partial \hat{k}^2$ can be read from the plot legends. The red dashed curve is the quantum critical
  line shown in Fig. \ref{PDtem0hee}.}\label{heeres1}
\end{figure}

Next we study the relation between the HEE and QPT in this Weyl
corrected EMA model. Notice that in scenario of HEE characterizing
QPT, the first order derivative to HEE \cite{Ling:2016wyr} or the
HEE itself \cite{Ling:2015dma} characterizes the QPT with local
extremes near QCPs. Nevertheless, in present model, the HEE itself
as well as its first derivative are featureless. Instead, we find
that the second order derivative of HEE with respect to system
parameter $\hat{k}$ exhibits peaks near the critical points of the
MIT. As we can see from Fig. \ref{line1}, $\partial^2 S/\partial
\hat k^2$ reaches its local extreme near the critical $k_c =
0.58116$, regardless of the value of $\gamma$.  To demonstrate
this phenomenon more transparently, we show the contour plot of
$\partial^2 S/\partial \hat k^2$ over the $(\gamma, \hat k)$ in
Fig. \ref{heeres1}. From this plot it is easily seen that the
local extremes (ridge for $\gamma<0$ and valley for $\gamma>0$) of
$\partial^2 S/\partial \hat{k}^2$ is close to the quantum critical
line. We would like to point out that the connection
between the $\partial^2 S/\partial \hat k^2$ and the critical line
is prominent only for $l>10^5$. The bigger the $l$ is, the better
the $\partial^2 S/\partial \hat k^2$ diagnoses the QPT.
This phenomenon is a natural result since the quantum critical
properties, which emerge at large scale, are expected to be
captured by large scale HEE.
Specially, in the limit of large $l$ we find that the $S/l$
converges to the Wald entropy density $s$, indicating that $s$ is
also a good indicator of the MIT in the present holographic model.
Similar phenomenon has been observed in \cite{Ling:2015dma} as
well. This phenomenon is in accordance with CMT result that the
entanglement measure characterizing the QPT becomes more prominent
with the increase of block size $l$.

Finally, it is noticed that there exists a mild discrepancy
between the ridge/valley and the quantum critical line.
One peculiar feature of our model
is that the Wald entropy density is non-zero for both metallic and
insulating phases even at zero temperature, thus the thermal
contribution to HEE can not be ignored and its effect to such a
discrepancy is unclear. At this stage we propose that mutual
information, which subtracts out the thermal contribution from the HEE,
might play a better role in diagnosing the QPT. We intend to study this in future.

\section{Discussions and open questions}\label{Con}

In this paper, we have constructed perturbative black brane
solutions to EMA-Weyl gravity model and studied the electrical
transport properties. A MIT is observed in our model, that is the
first realization of MIT in holographic models with higher
derivative gravity. We have also investigated the relation between
HEE and MIT, and found that the second order derivative of HEE
with respect to system parameter $\hat{k}$ exhibits peaks or
valleys near the critical points of MIT. Our results further
testifies the conjecture in \cite{Ling:2015dma,Ling:2016wyr} and
enriches the scenario of HEE characterizing QPTs. Certainly, it
can be expected that HEE characterizing QPT with even higher
orders of derivatives could be observed in holographic models.
These two results are of crucial importance to
a comprehensive understanding of quantum critical phenomena and the long-standing
question in CMT - why and how the EE characterize the QPT.

After a series of work on the relation between HEE and QPT,
it becomes urgent to understand the underlying reasons that lead
to different derivative orders of HEE diagnosing the QPT in
holographic approach, which is also an open problem in CMT.
Previously, it was argued in CMT literature for instance in
\cite{Larsson:2006sef} that the derivative order of entanglement
which becomes extremal or divergent might be related the order of
the QPT, which is determined by the behavior of free energy of the
system. Based on our results, however, this
correspondence is not observed in holographic framework.
Nevertheless, it is instructive to summarize and compare what we
have observed in this series of work.
\begin{enumerate}
  \item In \cite{Ling:2015dma}, HEE itself exhibits local extremes near the QCPs of the
  MIT. The ground state entropy density is \emph{vanishing} for
insulating phases, while \emph{nonvanishing} for the metallic
phase, reflecting an $AdS_2$ near horizon geometry.
  \item In \cite{Ling:2016wyr}, it is the first order derivative of HEE with respect to the system parameter
that diagnoses the QCPs of the MIT. In this circumstance both
metallic phase and insulating phase have \emph{vanishing} ground
state entropy density.
  \item In present paper, the second order derivative of HEE with respect to
the relevant parameter $\hat k$ characterizes the QPT.
Correspondingly, both metallic phase and insulating phase have
\emph{nonvanishing} ground state entropy density.
\end{enumerate}
Therefore we intend to propose that the derivative order of HEE
which signals QPT might be related to the behavior of ground
state entropy density in holographic approach. We also expect that
a well-designed quantity that removes the thermal contribution
from HEE, for instance the mutual information \cite{Kundu:2016dyk}, might play a
crucial role in unveiling the relation between the derivative
order of HEE and the QPT. We leave all these important issues for
further study. Next we point out some other interesting
topics worthy of further investigation.

First, it would be interesting to incorporate the holographic superconductor into our current framework.
In \cite{Wu:2010vr}, the Weyl corrected holographic superconductor without backreaction is constructed.
An important feature is that the ratio $\omega_g/T_c$ of gap frequency $\omega_g$ over critical temperature $T_c$ of superconducting phase transition
runs with the Weyl parameter $\gamma$. In particular, when $\gamma<0$, the value of $\omega_g/T_c$ is lower than that $\omega_g/T_c\simeq 8$
in the usual holographic superconductor \cite{Hartnoll:2008vx,Horowitz:2008bn,Hartnoll:2008kx}.
These results have been confirmed in subsequent series of works,
see for example  \cite{Ma:2011zze,Momeni:2011ca,Momeni:2012ab,Zhao:2012kp,Momeni:2013fma,Momeni:2014efa,Zhang:2015eea,Mansoori:2016zbp}.
It would be interesting to see how the ratio $\omega_g/T_c$ is affected by the $\gamma$ and $\hat k$ in our model.

Another worthwhile improvement to our current work is to obtain
black brane solutions with full backreaction, which involves
solving the differential equations beyond the second order.
Although our results are robust, but the first order approximation
requires the $\gamma$ to be very small. Solutions with full backreaction
will allow us to explore the properties of the system at wider range of $\gamma$.
When full backreaction is considered, the following important issues
could be addressed. First, our present perturbative EMA-Weyl
background has AdS$_2$ IR geometry at zero temperature, which
associates with a finite ground state entropy density. It would be
valuable to examine the behavior of the ground state entropy
density with the full backreaction. Second, the computation of
optical conductivity with full backreaction will reflect a more
accurate phase structure. Furthermore, we could study further
the mild discrepancy between the ridge of the HEE and the critical
line when the full backreaction is considered. In addition, it is
also interesting to include the superconductor in our present
model with full backreaction.

Finally, we would like to point out that it is sobering to realize that
our model may suffer from the micro-causality violation and the
instability of the dual CFT because of the introduction of the higher derivative term in
action (\ref{action}) \cite{Ritz:2008kh,Brigante:2007nu,Brigante:2008gz}.
By analyzing the causality of the CFT dual to the Schwarzschild-AdS geometry
and the stabilities of the vector modes, a constraint is placed
on $\gamma$ \cite{Myers:2010pk,Ritz:2008kh}. While the backreaction of
higher derivative term is included, the causality and the stability
should be reexamined. In this situation, however, we need to turn on not only vector modes but
also scalar and tensor modes to analyze their causality and
stability, and the resulting equations are much more
complicated than \cite{Myers:2010pk,Ritz:2008kh}. At this stage, we leave the rigorous analysis
for future investigation.

\section*{Acknowledgements}

We are very grateful to Wei-Jia Li for helpful discussion.
This work is supported by the Natural Science Foundation of China under
Grant Nos.11275208, 11305018 and 11575195, and by the grant (No.
14DZ2260700) from the Opening Project of Shanghai Key Laboratory
of High Temperature Superconductors. Y.L. also acknowledges the
support from Jiangxi young scientists (JingGang Star) program and
555 talent project of Jiangxi Province. J. P. Wu is also supported
by the Program for Liaoning Excellent Talents in University (No. LJQ2014123).

\end{document}